\documentclass[12pt,epsfig]{article}
\usepackage{graphicx}

\newcommand{\be}[1]{\begin{equation} \label{#1}}
\newcommand{\ee}{\end{equation}}

\newcommand{\bea}[1]{\begin{eqnarray} \label{#1}}
\newcommand{\eea}{\end{eqnarray}}

\newcommand{\refeq}[1]{(\ref{#1})}

\begin{document}

\begin{center}
\Large \bf
         Non-Ergodic Nuclear Depolarization in Nano-Cavities.
\end{center}
\begin{center}
\large \bf
               E.B. Fel'dman and M.G. Rudavets
\end{center}
\begin{center}

       Institute of Problems of Chemical Physics, \\
       Russian Academy of Sciences, 142432 Chernogolovka,
       Moscow Region, Russia
\end{center}


\begin{abstract}

Recently, it has been observed that the effective dipolar
interactions between nuclear spins of spin-carrying molecules of a
gas in a closed nano-cavities are independent of the spacing
between all spins. We derive exact time-dependent polarization for
all spins in spin-$\frac{1}{2}$ ensemble with spatially independent
effective dipolar interactions. If the initial polarization is on a
single (first) spin,$P_1(0)= 1$ then the exact spin dynamics of the
model is shown to exhibit a periodical short pulses of the
polarization of the first spin, the effect being typical of the
systems having a large number, $N$, of spins. If $N \gg 1$, then
within the period $4\pi/g$ ($2\pi/g$) for odd (even) $N$-spin
clusters, with $g$ standing for spin coupling, the polarization of
spin $1$ switches quickly from unity to the time independent value,
$1/3$, over the time interval about $(g\sqrt{N})^{-1}$, thus,
almost all the time, the spin $1$ spends in the time independent
condition $P_1(t)= 1/3$.
The period and the width of the pulses determine
the volume and the form-factor of the ellipsoidal cavity. The
formalism is adopted to the case of time varying nano-fluctuations
of the volume of the cavitation nano-bubbles. If the volume $V(t)$ is
varied by the Gaussian-in-time random noise then the envelope of
the polarization peaks goes irreversibly to $1/3$.
The polarization dynamics of the single spin exhibits the Gaussian
( or exponential )
time dependence when the correlation time of the fluctuations
of the nano-volume is larger ( or smaller )
than the
$
\langle ( \delta g )^2 \rangle^{-1/2}
$, where the
$
\langle ( \delta g )^2 \rangle
$
is the variance of the $g(V(t))$ coupling.
Finally, we report the exact
calculations of the NMR line shape for the $N$-spin gaseous aggregate.

\end{abstract}
\nopagebreak


\newpage
\noindent
\section{Introduction}
\label{sec. 0}

The nature of ergodicity being of fundamental importance for
consonant description of statistical mechanics is currently being
discussed in the NMR context \cite{Jheparov}. Spin dynamics is ergodic if the
initial polarization prepared at a single (first) spin is spread
over the system leading, as time proceeds, to the spatially uniform
distribution of the polarization, as it is expected on the basis of
a simple physical intuition. Instead, non-ergodic behavior that
have recently been observed numerically in the nuclear
spin-$\frac{1}{2}$ $1D$ chains under the general XYZ spin
Hamiltonian \cite{BE} enters in the way that the time average
polarization of the first spin turns out to be several times larger
as compared to that of any other spin in the chain. This
observation of nonergodicity has been extended to $1D$ chains and
rings under XY Hamiltonian \cite{FBE} showing analytically that the
time average polarization of the first spin differs by the factor
$1.5 \div 2$ from the time average polarization of all other spins
in the chain. These considerations in $1D$ spin clusters address
the problem of the nature of the ergodicity to  different spin
Hamiltonians. Motivated in studying of non-ergodic spin dynamics
and due to the fact that exact solution is the lucky exception in
statistical mechanics, our assumption in this paper is that spin
interactions are regarded to be independent of the spacing between
the spins rather than having $r^{-3}$ dependence.

Recently, the spin Hamiltonian with space-independent spin couplings
has been applied for exploring the NMR spectra of a gas of
spin-caring molecules undergoing a fast thermal motion within the
non-spherical cavities \cite{Baugh}. In that report, the authors
have arrived at the space-independent effective spin couplings by
motionally average the exact dipolar Hamiltonian over a uniformly
distributed spins' spatial coordinates in a nanometer sized
cavities. This technique is expected to have a promising
application for determining the pores' shapes and sizes \cite{Inag}
by the NMR spectra.

With regard to the effective nuclear spin Hamiltonian with
infinite range couplings it is noteworthy that this type of interactions has
also been proposed in the theory of nano-electrodes \cite{Kane},
\cite{Moz}.
Infinite range dipolar nuclear interactions are
induced indirectly due to the fast energy transfer between electron and nuclear
spins.
On the coarse grained time scale of the fast electron spin dynamics, the slow
nuclear spin dynamics is governed by an effective nuclear spin
Hamiltonian with an infinite range interaction coupling. Quite apart
from its importance as a physical model in the NMR experiments for
a many-spin aggregate in a confined volume \cite{Baugh}, \cite{Kane}, \cite{Moz} and
few proton molecules \cite{Yan}, the model with infinite range
spin interactions is of a fundamental interest in its own right since
this model allows to treat 3-dimensional case exactly without any
reference to an $1D$ spin ordering. It represents the quantum
non-equilibrium version of the exactly solvable equilibrium spin
model \cite{KH}, has the mapping to the BCS pairing Hamiltonian of
the superconductivity \cite{AW}, and has long been provided to be
the test for many-body problems in high spatial dimensions, $D \gg
1$. The objective of the paper is to present the exact solution of
non-ergodic dynamics with infinite range spin Hamiltonian in $N$-
spin-$\frac{1}{2}$ ensemble.

To our knowledge, the only result reported on this model is that of
Waugh \cite{W} who announced (without proof) that the time average
polarization of the first spin equals exactly $\frac{N+2}{3N}$ and
the polarization of any other spin is exactly $\frac{2}{3N}$ for
odd numbered, $N$, spin cluster. To clarify the problem of spin
dynamics, the present paper reports a detailed analytical theory of
the average polarization both for odd and even numbered spin
clusters as well as it gives the theory of spin dynamics that is
entirely missing from \cite{W}. A condensed form of this paper have
been published in \cite{RF}. Brief overview of the paper is as
following. In section $2$, we construct the effective nuclear spin
Hamiltonian of spin-carrying molecules of the perfect gas in the
nano-cavity. Section $3$ gives formalism required to obtain the
exact time-dependent polarization. This is followed by Section $4$
that discusses three issues of the polarization dynamics that
amenable to the techniques of the Section $2$: firstly, the
non-ergodicity of the polarization dynamics of a single spin in the
nano-cavity; secondly, the polarization dynamics of a single spin
within a fluctuating nano-bubbles; thirdly, spectral line shape of
the nuclear spin ensemble. Finally, Section $5$ summarizes the
conclusions of the calculations and confronts the results obtained
with known analytical results for the XY Hamiltonian.


\noindent
\section{ Effective Nuclear Spin Hamiltonian in Nano-Cavity }
\label{sec. 1}

The purpose of this section is to construct an effective spin
Hamiltonian, $H_{\sf eff}$, that governs spin dynamics of
spin-carrying molecules in a nano-sized cavity on the coarse
grained
temporary scale of the order $10$ picoseconds. On these space-time scales
the effective spin Hamiltonian differs from the exact dipolar
Hamiltonian, in particular, the $H_{\sf eff}$ has a high symmetry
that permits the exact solution for the spectrum of the many-body
spin Hamiltonian $H_{\sf eff}$ and, as a result, the exact
derivation of the polarization dynamics of the gas within the
nano-cavity. In this section we summarize the main ideas of the
report \cite{Baugh}, however in derivation of the effective spin
Hamiltonian $H_{\sf eff}$ by averaging over spins' spatial
coordinates, we generalize the effective spin coupling to the case
of non-perfect gas in the nano-cavity.

The starting point of the derivation of the operator
$H_{\sf eff}$ is the expression for the density matrix
\bea{1.1}
\rho(t,\{\vec I_n,
\vec r_n(t), \vec p_n(t) \}_{n=1}^{N} ) =
U(t)
\rho(0,\{\vec I_n,
\vec r_n(0), \vec p_n(0) \}_{n=1}^{N} )
U^{-1}(t)
\eea
for completely specified coordinates
$\{ \vec r_n(t) \}_{n=1}^{N}$
and momenta
$\{ \vec p_n(t)\}_{n=1}^{N}$
of $N$ spin-carrying molecules. The propagator $U(t)$
is associated with the time dependent exact dipolar Hamiltonian (in frequency units)
\bea{1.2}
 H(t) =
  \sum \limits_{ 1  \le  i  <  j }^{N} h_{i,j}(t), \quad
  h_{i,j}(t) =
 \gamma^2 \hbar
         P_2(\cos\theta_{ij}(t))
          r_{ij}^{-3}(t)
         ( \vec I_{i} \vec I_{j} - 3I_{iz} I_{jz} ),
\eea
where, $\gamma$ stands for the gyromagnetic ratio, $I_{n\alpha}
(\alpha = x,y,z)$ specify the spin-$\frac{1}{2}$ operators, the
$\theta_{ij}(t)$ is the instant polar angle between the vector $r_{ij}(t)$
from $r_i(t)$ to $r_j(t)$ and the external magnetic field $B$.

A cornerstone fact for a construction of an effective spin
Hamiltonian is an essential difference between the time scale of
the relaxation in the phase space $r^N - p^N$ and the time scale of
the spin dynamics under the Hamiltonian of Eq. \refeq{1.2}.
Actually, for the hydrogen gas at room temperature and atmospheric
pressure, the following estimations hold. The average concentration
$\bar n \approx 2.7 \cdot 10^{19}$ molecules/cm$^3$, the mean free path
$\lambda = (\bar n \pi a^2)^{-1} \approx 10^{-4}$ cm for the radius
of a molecule $a \approx 10^{-8}$ cm and the thermal velocity
$\bar v \approx 10^{5}$ cm/s.
Then, a simple
order-of-magnitude calculations leads us to expect that, for the
gas in the cavity of the size $\ell \approx 10$ nm,
the diffusion coefficient
${\cal D} \approx \bar v \ell  \approx 10^{-1}$ cm$^2$/s, the
characteristic time scale of the spatial relaxation of the gas is
$t_{\sf dif} \approx \ell^2/{\cal D} \approx 10^{-11}$ s and the
characteristic time scale of the momenta relaxation towards the
Maxwellian distribution is
$t_{\sf v} \approx \lambda/{\bar v} \approx 10^{-11}$ s.
These time scales $t_{\sf v}$ and
$t_{\sf dif}$ are well separated from the NMR time scale
$t_{\sf nmr} = 10^{-4} \div 10^{-3}$ s associated with the dipolar interaction in Eq.
\refeq{1.2}. The smallness of the parameter
\bea{1.3}
\varepsilon =
\frac{ t_{\sf rel} }{ t_{\sf nmr} } =
10^{-7}  \ll 1, \quad \mbox{where} \quad t_{\sf rel} = \max( t_{\sf v},t_{\sf dif} )
\eea
allows for determining the average nuclear spin Hamiltonian
governing the behavior of the nuclear spins over a
coarse-grained time intervals $\Delta t$
obeying
\bea{1.4}
t_{\sf rel} \ll \Delta t \ll  t_{\sf nmr}.
\eea
The averaging  of the exact Hamiltonian over time
$\Delta t$
is
performed to the zeroth order in the perturbation expansion in powers
of the parameter $\varepsilon$  yielding the average
(or effective) Hamiltonian \cite{Haeberlen},
\bea{1.5}
\overline{ h_{i,j} }  =
 \frac{1}{ \Delta t  }\int_0^{ \Delta t  }
 h_{i,j}(  r_i(t'), r_j(t')  ) dt' ,
\eea
with the corrections being of the order ${\cal O}( \varepsilon^{1} )$.

The decisive point of the following treatment is the replacement
of the temporary integration  in Eq. \refeq{1.5} with
the integration over the spatial coordinates within the confined region.
Equating the temporary averaging with the spatial averaging makes
sense under the ergodic hypotheses \cite{Khinchin}:
\bea{1.6}
\frac{ \delta t( d\,r^N ,  d\,p^N ) }{ t }
=
Z^{-1} e^{-E/kT} d\,r^N  d\,p^N   ,
\eea
where the notations imply that a representative point living in the
whole phase space $r^N - p^N$, while moving over the time $t$,
$t_{\sf rel} \ll t \ll t_{\sf nmr} $, spends
within the volume $d\,r^N d\,p^N$
only a fraction
$Z^{-1} e^{-E/kT} d\,r^N d\,p^N$ of the whole time $t$,
the $E$ being the total energy.
The relation \refeq{1.6} incorporates the
Gibbsian stochastic level of description into the dynamical
treatment of the molecular collisions. The phase
space degrees of freedom are assumed to relax to their equilibrium
distribution at a given temperature $T$.

Introduce equilibrium pair distribution function for molecules $1$ and $2$,
\bea{1.7}
D_2(r_1, r_2) =
\frac
{\int_V   \, d^3r_3 \cdots \int_V d^3r_N
\exp(-U(r^N)/kT)}
{\int_V   \, d^3r_1 \cdots \int_V d^3r_N
\exp(-U(r^N)/kT)},
\eea
where $U(r^N)$ stands for inter-molecular electrostatic
interactions, (recall that
$
\| U \|/ \| H \| \approx  10^{7}
$ \cite{Abragam}).
Then, taking for granted the ergodicity \refeq{1.6}, the
evolution of spin degrees of freedom on the coarse grained time scale
$\Delta t$ of Eq. \refeq{1.4} is governed by the static (time
independent) effective Hamiltonian
\bea{1.8}
\overline{ H } =
\sum \limits_{ 1  \le  i  <  j }^{N} \overline{ h_{i,j} },
 \qquad
\overline{ h_{i,j} } =
 g \sum \limits_{ 1  \le  i  <  j }^{N}
         ( \vec I_{i} \vec I_{j} - 3I_{iz} I_{jz} )
\eea
with spacing independent pair couplings $g$ for any pair of
spins $i$ and $j$,
\bea{1.9}
g =
 \gamma^2 \hbar
 \int_V \int_V  d^3r_i \, d^3r_j D_2 ( r_i, r_j )
          P_2(\cos\theta_{ij})
          r_{ij}^{-3}  .
\eea
The effective operator $\overline{ H } $
involves only the (slow)
spin operators, whereas the (fast) spatial coordinates of the
nucleus ( labeled by indices $i$ and $j$) are integrated out.
On the coarse grained scale $\Delta t$, any nuclear spin
"feels" the field that is independent of the spatial coordinates
of all other spins flying within the nano-cavity
but depend on the quantum states of that
spins.

The effective spin
coupling $g$ encodes an information about the shape and the size of the nano-cavity.
The primary objective of the proceeding discussion is to present
the expression for the coupling $g$ of Eq. \refeq{1.9}
for an ellipsoidal nano-cavity.

For a perfect gas within the nanocavity, the pair distribution function equals
$D_2 ( r_i, r_j )= V^{-2}$
for spins $i$ and $j$ lying in the cavity $r_i, r_j \in V$, hence
the averaging in Eq. \refeq{1.9} gives
\bea{1.21}
g =
\gamma^2 \hbar F/V ,  \qquad
F =
\int_V
\int_V  \frac{d^3 r_1 d^3 r_2}{V}
          P_2(\cos\theta_{12})
          r_{12}^{-3} .
\eea
In this report, use is made of the remarkable fact \cite{Baugh}
that the  volume, $V$, of the nano-cavity enters into the
expression for the effective coupling $g$ of Eq. \refeq{1.21},
which itself enters into the polarization (defined below in the
section \ref{sec. 2}) giving rise to the dependence of the
polarization on the volume of the nano-cavity by no means as
trivial as merely proportional to the volume.

The transformation of the coordinates $r_1$ and $r_2$
to the relative coordinate $r_{12} = r_1 - r_2$
and the coordinate of the center of
gravity, $r = (r_1 + r_2)/2$,
reduces the form-factor $F$ in Eq. \refeq{1.21} to the form
\bea{1.21.2}
F = \int_V  d^3r_{12} \,
        P_2(\cos\theta_{12})
        r_{12}^{-3} .
\eea
It is convenient to assume that the initial point of the vector
$r_{12} = r_1 - r_2$ starts at the origin of the frame of
reference $xyz$ connected with the ellipsoid, see Figure  $1$.
In order to calculate the integral in Eq. \refeq{1.21.2} in the ellipsoid
frame of reference, we define by
    $\theta_{\mathbf{z},\mathbf{r}_{12} }$
and $\phi_{\mathbf{z},\mathbf{r}_{12} }$ the polar and the azimuthal angles of
the vector $r_{12}$ in the ellipsoid frame of reference $xyz$,
respectively, and the angles $\theta_{ \mathbf{Z},\mathbf{r}_{12} }$,
$\phi_{ \mathbf{Z},\mathbf{r}_{12} }$ will
be referred to as the polar and the azimuthal angles for the vector
$r_{12}$ in the laboratory frame of reference $XYZ$. By the
"addition" theorem of spherical harmonics, we can write for the
factor
$
P_2( \cos\theta_{12} )
$ in Eq. \refeq{1.21.2}
the expression
$
P_2( \cos\theta_{12} )=
P_2(\cos\theta_{\mathbf{Z},\mathbf{r}_{12}})=
P_2(\cos\theta_{
\mathbf{z},\mathbf{Z} })
P_2(\cos\theta_{\mathbf{z},\mathbf{r}_{12}}) +
\ldots
$
where the dots denote the terms depending  on the azimuthal angles as
$
\exp\{ im (\phi_{\mathbf{z},\mathbf{r}_{12}}
- \phi_{ \mathbf{z},\mathbf{Z} } )   \}
$
with
$m \ne 0$,
hence, all these terms  vanish after integration
over the angle
$
\phi_{\mathbf{z},\mathbf{r}_{12}}
$.
Defining shorthand
$
\phi = \phi_{\mathbf{z},\mathbf{r}_{12}}
$,
$
\theta = \theta_{\mathbf{z},\mathbf{r}_{12}}
$,
$
\alpha = \theta_{ \mathbf{z},\mathbf{Z} }
$, we get
\bea{1.22}
F = P_2(\cos\alpha) F', \quad
F' =  \int_0^{2\pi} d \phi
           \int_0^{\pi} \sin \theta d\theta
            P_2(\cos\theta)
           \int_0^{ R(\theta)}
           r_{12}^{-1} dr_{12} .
\eea
The Eq. \refeq{1.22} shows that the dependence of the form-factor on the angle
$\alpha$
is factored out. The remaining integral, $F'$, in Eq. \refeq{1.22}
coincides exactly with the form-factor of the ellipsoid
having the $z$ axis along  the $Z$ axis of the laboratory frame of reference.
The radius
$R(\theta)$
stands for the surface of the ellipsoidal cavity.
To find the integral in Eq. \refeq{1.22}, we parameterize  the ellipsoid
$
(z/a)^2 + (x^2 + y^2)/b^2 = 1
$
with the spherical coordinates
$\rho$ and $\theta$ as follows,
$
\rho = b(1 - \varepsilon^2 \cos^2\theta)^{-1/2}
$,
where
$
\varepsilon^2 = 1 - (b/a)^2
$.
In the spherical coordinates, the form-factor $F'$ becomes
\bea{1.23}
F' =   \int_0^{2\pi} d \phi
           \int_0^{\pi} \sin \theta d\theta
            P_2(\cos\theta)
           \ln \left. \left[
               \frac{ b/R_{\sf min} }
                    { (1 - \varepsilon^2 \cos^2\theta )^{1/2}  }
                 \right]
                 \right|_{ R_{\sf min} \to 0 },
\eea
here, the auxiliary lower limit
$
   R_{\sf min} \to 0
$
is introduced for the convergence of the integral. The constant
$R_{\sf min}$ vanishes from the final form-factor $F'$. Indeed,
invoking the definition $P_2(\cos\theta) = (3 \cos^2\theta - 1 )/2$
and using the integral
$
\int\nolimits_{-1}^{1}P_2(x) dx = 0
$, we get rid of the dependence on $R_{\sf min}$ in the integral
$F'$ of Eq. \refeq{1.23}, thus, reducing the integral $F'$ to the form
\bea{1.24}
F' =  - \frac{1}{2} \int_0^{2\pi} d \phi
            \int_0^{\pi} \sin \theta d\theta
            P_2(\cos\theta)
           \ln ( 1 - \varepsilon^2 \cos^2\theta ) .
\eea
Straightforward integrating over $\theta$ by parts and combining
the result with Eq. \refeq{1.22} yields the sought for form-factor
\bea{1.24}
F =  I \pi P_2(\cos\alpha),  \quad  I= \cases{
 \frac{2}{3} + 2 \left(  \frac{1}{\varepsilon^2} - 1  \right)
                 \left( 1 -  \frac{1}{\varepsilon} \mathop{\rm arcth} \varepsilon\right)
    , & $\mbox{for $a \ge b$} $ \cr
  \frac{2}{3} - 2 \left( \frac{1}{|\varepsilon |^2} + 1 \right)
                 \left( 1 -  \frac{1}{|\varepsilon|} \mathop{\rm arctg} |\varepsilon | \right)
    , & $ \mbox{for $a \le b$ .} $ \cr
                                            }
 \eea
For $a \gg b $, the $\varepsilon = 1 $ and $I = 2/3$.
For $a = b $, the $\varepsilon \to 0 $ and
Eq. \refeq{1.24} gives $I = 0$.
For $a \ll b $, the $ | \varepsilon | \to \infty $ and
$I = \frac{2}{3} - 2 = -4/3$. These limiting cases confirm the result reported in \cite{Baugh}.


\noindent
\section{Polarization  Dynamics}
\label{sec. 2}

We regard the spin Hamiltonian $\overline{ H }$
of a $N$-spin cluster in a uniform external magnetic field $B$ parallel to
the Z-axis of a fixed frame of reference $XYZ$ and the spatially independent spin couplings $g$ of Eq. \refeq{1.21},
\bea{2.1}
\overline H  =
\omega \sum\limits_{n=1}^{N} I_{nz}
+
\frac{g}{2} \sum\limits_{m \ne n}^{N}
\left\{
\zeta I_{mz} I_{nz} - I_{mx} I_{nx} - I_{my} I_{ny}
\right\},
\eea
here, $\omega = \gamma B$ denotes the Zeeman frequency and
the $\zeta$ is an arbitrary factor.

The standard way of approaching $N$-spin aggregate is to find
the  polarization
on the $n$-th spin at a time instant $t$ given initial polarization
on the $1$-st spin
\bea{2.2}
P_n(t) =
tr
\left\{
 e^{i\overline{ H }t} I_{1z} e^{-i\overline{ H }t} I_{nz}
\right\}
/
tr
\left\{
I_{1z}^2
\right\} .
\eea
The Hamiltonian of Eq. \refeq{2.1} can be cast
into the form (up to the constant  $gN(1 - \frac{\zeta}{2} )/4$  )
\bea{2.1.1}
\overline{ H } =
\omega I_{z}
+
\frac{g}{2}( \zeta + 1) I_{z}^2 -
\frac{g}{2} {I}^2   ,
\eea
where the total spin
$ I = \sum\nolimits_{n=1}^{N} I_n$
and its projection onto the z-axis reads
$ I_z = \sum\nolimits_{n=1}^{N} I_{nz}$.
The polarization
$P_n(t)$  of Eq. \refeq{2.2} is unchanged if we modify
the
Hamiltonian $\overline{ H }$ of Eq. \refeq{2.1.1} to the effective one
\bea{2.3}
H' = - \frac{g}{2}  I^2 .
\eea
In addition, the equivalence of polarizations $P_n(t)$
of all spins, except the first spin, as well as
the conservation with the time of the total polarization
$ \sum\nolimits_{n=1}^{N} P_n(t) $
allow to concentrate the description on the first spin only,
\bea{2.4}
P_1(\tau) =
tr
\left\{
 e^{i\tau  I^2 } I_{1z} e^{-i\tau  I^2 } I_{1z}
\right\}
/
tr
\left\{
I_{1z}^2
\right\},
\eea
here, the dimensionless time scale is defined according to
$\tau = \frac{1}{2}gt$.
A powerful tool for investigating the problem of Eq. \refeq{2.4}
is the theory of coupling of the angular momenta \cite{LL}, \cite{Mes1}.
To this end, think of the total spin cluster as composed of two
subsystems $A$ and $B$. The subsystem $A$ has only
the spin $ I_1 =  I_A$
and the remaining fragment $B$ of the spin cluster
has the spin $ I_B$, with the total spin being
$ I = I_A + I_B$.
The states of the two subsystems  $A$ and $B$
are coupled together within the state of the whole system $A\oplus B$
through the
Clebsch-Gordan (C-G) coefficients,
\bea{2.5}
|I_A, I_B, I, m \rangle =
 \sum\limits_{m_A = \pm \frac{1}{2}  \atop   m_B = m - m_A  }
C_{I_A, m_A; I_B, m_B }^{ I, m }
|I_A m_A \rangle
|I_B m_B  \rangle ,
\eea
where
$I_A =\frac{1}{2}$ and $ m_A = \pm \frac{1}{2}$
are the spin and its magnetic quantum numbers for the 1-st spin,
respectively, and
$I_B$ and $m_B = m - m_A$ stand for
spin and its magnetic quantum numbers of the fragment $B$.
For $I_B =0$ only $I=\frac{1}{2}$ and $m = \pm \frac{1}{2}$ are allowed.
For $I_B \ge \frac{1}{2}$, the allowed $I$ and $m$ are as follows,
$I = I_B \pm \frac{1}{2}$, $-I \le  m  \le I$.
The C-G coefficients are, see e.g. \cite{Bethe},
\bea{2.5.01}
C_{\frac{1}{2}, \frac{1}{2}; I_B, m - \frac{1}{2}}^{ I_B + \frac{1}{2}, m }
&=&
C_{\frac{1}{2}, -\frac{1}{2}; I_B, m + \frac{1}{2}}^{ I_B - \frac{1}{2}, m }
=
[ ( I_B + \frac{1}{2} + m )/(2 I_B + 1) ]^{1/2} ,
\nonumber  \\
C_{\frac{1}{2}, -\frac{1}{2}; I_B, m + \frac{1}{2}}^{ I_B + \frac{1}{2}, m }
&=&
C_{\frac{1}{2}, \frac{1}{2}; I_B, m - \frac{1}{2}}^{ I_B - \frac{1}{2}, m }
=
[ ( I_B + \frac{1}{2} - m )/(2 I_B + 1) ]^{1/2} .
\eea
The two pairs of independent variables $(I_B, m_B)$ and $(I_A = \frac{1}{2}, m_A = \pm \frac{1}{2})$
will be used for determining the trace in Eq. \refeq{2.4} for the whole $N$ spin
system $A\oplus B$,
\bea{2.5.1}
tr
\left\{ \dots \right\} =
      \sum\limits_{ I_B = I_B^{\sf min} }^{N_B/2} w(I_B)
      \sum\limits_{ I = | I_B - \frac{1}{2} | }^{ I_B + \frac{1}{2} }
      \sum\limits_{ m = -I }^{ I }
\langle I_A, I_B, I, m |
\dots
|I_A, I_B, I, m \rangle ,
\eea
where $N_B = N-1$ is the number of spins in the fragment $B$, the
minimal value of $I_B$ is $I_B^{\sf min} = 0$ for even $N_B$ and
$I_B^{\sf min} = \frac{1}{2}$ for odd $N_B$.
The factor
\bea{2.5.2}
w(I_B) = \frac{2I_B + 1}{N_B + 1}
{  N_B + 1  \choose \frac{1}{2}N_B + I_B + 1 }
\eea
is the number of ways of grouping $N_B$ individual
spin-$\frac{1}{2}$
into the total spin $I_B$. The $w(I_B)$
obeys the property  \cite{LL}, \cite{Mes1}, \cite{Mes2}
\bea{2.5.3}
\sum\limits_{I_B \ge | m_B |}^{N_B/2} w(I_B) =
{  N_B  \choose \frac{1}{2}N_B + m_B }.
\eea
The right-hand side of Eq. \refeq{2.5.3} is the number of states for
each allowed eigenvalue $m_B$ of the fragment B.
In order to deal with the diagonal evolution  matrices within the
Eq. \refeq{2.4},
we introduce additional basis of the bra,
$\langle I_A, I_B, I',m' |$,
and the ket, $ |I_A, I_B, I', m' \rangle$,
vectors of the Hilbert space ${\cal H}(I_A)\otimes {\cal H}(I_B)$ for
fixed values $I_A = 1/2$ and the $I_B$, then make use of the
completeness of the $2(2I_B + 1)$ orthonormal basis vectors
belonging to the space ${\cal H}(I_A)\otimes {\cal H}(I_B)$
\bea{2.5.4}
1_{{\cal H}(I_A)\otimes {\cal H}(I_B)} =
      \sum\limits_{ I' = | I_B - \frac{1}{2} | }^{ I_B + \frac{1}{2} }
      \sum\limits_{ m' = -I }^{ I }
|I_A, I_B, I', m' \rangle
\langle I_A, I_B, I', m' | ,
\eea
and, finally, insert the representation of the unity of Eq. \refeq{2.5.4}
in front of the right-most operator $I_{1z}$ of Eq. \refeq{2.4} having the
matrix elements
\bea{2.5.5}
\langle I_A, I_B, I', m' |
I_{1z}
|I_A, I_B, I, m \rangle = \delta_{m,m'}
      \sum\limits_{ m_A = \pm \frac{1}{2} }
       m_A  C_{\frac{1}{2}, m_A; I_B, m'- m_A }^{ I', m' }
            C_{\frac{1}{2}, m_A; I_B, m - m_A }^{ I , m  } .
\eea
With these algebraic steps, we immediately get the
polarization $P_1(\tau)$
in terms of the Clebsch-Gordan coefficients as
\bea{2.6}
\lefteqn{ P_1(\tau) =
              2^{-(N_B-1)}
              \sum\limits_{ I_B = I_B^{\sf min} }^{N_B/2}
               w(I_B)
\sum\limits_{  | I_B - \frac{1}{2} | \le I \le   I_B + \frac{1}{2}
         \atop | I_B - \frac{1}{2} | \le I' \le  I_B + \frac{1}{2}
            }
\sum\limits_{   -I \le  m \le  I
         \atop  -I' \le  m' \le  I'
            }    \delta_{m, m'}  \times
        }                                    \nonumber      \\
&                  &
 \times   e^{ i\tau\{I(I+1) - I'(I'+1)\} }
 \Bigl(
      \sum\limits_{ m_A = \pm \frac{1}{2} }
       m_A  C_{\frac{1}{2}, m_A; I_B, m'-m_A }^{ I', m' }
            C_{\frac{1}{2}, m_A; I_B, m-m_A }^{ I , m  }
\Bigr)^2 .
\eea
For the term
$I_B = 0$, only the single pair
$(I=\frac{1}{2}, I'=\frac{1}{2})$
is allowed in the sum Eq. \refeq{2.6}, and for $I_B \ge
\frac{1}{2}$,
the four pairs of $(I,I')$ should be distinguished in the sum Eq.
\refeq{2.6} depending on the sign $(+)$ or $(-)$ in the
expressions
\bea{2.6.1}
(I,I') = \left( I =  I_B \pm \frac{1}{2}, \,  I' = I_B \pm \frac{1}{2}
         \right) .
\eea

Armed with the polarization  $P_1(\tau)$ of Eq. \refeq{2.6},
we are going to decompose
it into the time-independent part $\overline{P_1}$ and the
oscillating part $P_1^{\sf osc}(\tau)$,
\bea{2.9}
P_1(\tau) = \overline{P_1} + P_1^{\sf osc}(\tau).
\eea
The time-independent contribution $\overline{P_1}$ to the function
$P_1(\tau)$ is provided by the quantum numbers $m, m'$ belonging to the
states
$I = I' = I_B \pm \frac{1}{2}$
if $I_B \ge \frac{1}{2}$, and by the quantum numbers $m, m'$ belonging to the states
$I = I' = \frac{1}{2}$
if $I_B = 0$,
\bea{2.10}
\overline{P_1} =
      2^{-(N_B-1)}
      \sum\limits_{ I_B = I_B^{\sf min} }^{N_B/2} w(I_B)
      \sum\limits_{ I = | I_B - \frac{1}{2} | }^{ I_B + \frac{1}{2} }
      \sum\limits_{ m = -I }^{ I }
\Bigl(
      \sum\limits_{ m_A = \pm \frac{1}{2} }
       m_A
\bigl(
       C_{\frac{1}{2}, m_A; I_B, m-m_A }^{ I, m  }
\bigr)^2
\Bigr)^2 .
\eea
Our desire now is to sum in Eq. \refeq{2.10} over the indexes $m$ and
$I$ for the fixed value of the $I_B$.
To this end, we start with the state $I_B=0$
that arises
for even $N_B$ ( see the comments to Eq. \refeq{2.5.1} ).
For $I_B=0$ only $I=\frac{1}{2}$ is allowed,
and the partial polarization
$\overline{P_1}(I_B)$ in Eq. \refeq{2.10} reads
\bea{2.11}
\overline{P_1}(I_B=0) =
      2^{-(N_B-1)} w(0)
      \sum\limits_{ m = - \frac{1}{2} }^{ \frac{1}{2} } m^2 .
\eea
Next, we consider the contribution to the $\overline{P_1}$ of Eq. \refeq{2.10}
from the spin $I_B \ge \frac{1}{2}$. In this situation,
$I =  I_B \pm \frac{1}{2} $ are allowed and invoking the
C-G coefficients of Eq. \refeq{2.5.01},
the contributions
$\overline{P_1}(I_B)$ to the $\overline{P_1}$ of Eq. \refeq{2.10}
can be written in a convenient form as
\bea{2.12}
\overline{P_1}(I_B) =
      2^{-N_B} w(I_B)
      \sum\limits_{ \mu = - I_B }^{ I_B }
      \frac{2\mu + 1}{2I_B + 1}   .
\eea
Combining together the $\overline{P_1}(I_B=0)$ of Eq. \refeq{2.11}
and  the $\overline{P_1}(I_B)$ of Eq. \refeq{2.12} results in
\bea{2.13}
\overline{P_1} =
      2^{-N_B}
      \sum\limits_{ I_B = I_B^{\sf min} }^{N_B/2} w(I_B)
      \sum\limits_{ \mu = - I_B }^{ I_B }
      \frac{2\mu + 1}{2I_B + 1}    .
\eea
The sum over $\mu$ in Eq. \refeq{2.13} yields easily
\bea{2.14}
\sum\limits_{ \mu = -I_B}^{I_B}
(2\mu + 1)^2
 = (2I_B + 1)( 1 + \frac{4}{3}I_B(I_B + 1) ) ,
\eea
and after substitution the value $w(I_B)$ of Eq. \refeq{2.5.2},
we arrive at sought for result
\bea{2.14.1}
\overline{P_1} =
\frac{ 2^{-N_B} }{N_B + 1}
\sum\limits_{ I_B = I_B^{\sf min} }^{N_B/2}
            {N_B + 1  \choose \frac{1}{2}N_B + I_B + 1 }
                              ( 1 + \frac{4}{3}I_B(I_B + 1) ).
\eea
The remaining sum over $I_B$ in Eq. \refeq{2.14.1} depends
on whether $N_B$ is an even or odd number.
If $N_B$ is an even number, then $I_B^{\sf min} =0$ and
straightforward summing over $I_B$
in $\overline{P_1}$ of Eq. \refeq{2.14.1} by exploiting the
following sums of the binomial coefficients,
\bea{2.14.2}
\sum\limits_{ I_B = 0}^{N_B/2}
            {N_B + 1  \choose \frac{1}{2}N_B + I_B + 1 } & = &
            2^{N_B}, \nonumber \\
\sum\limits_{ I_B = 0}^{N_B/2} I_B(I_B + 1)
            {N_B + 1  \choose \frac{1}{2}N_B + I_B + 1 } & = &
            N_B 2^{N_B-2}
\eea
yields the polarization
\bea{2.15}
\overline{P_1} = \frac{N+2}{3N}
\eea
for an odd $N = N_B + 1$ spin cluster \cite{W}.
If $N_B$ is an odd number,
then $I_B^{\sf min} = \frac{1}{2}$ and some simple algebra gives
the polarization
\bea{2.16}
\overline{P_1}  =
\frac{ N + 2 - 2^{1-N} { N \choose N/2 } }{3N}
\eea
for an even $N = N_B + 1$ spin cluster.

When $N \gg 1 $, the $\overline{P_1}$
of Eq.  \refeq{2.16}
behaves as $(N + 2 - 2 \bigl( \pi N/2)^{-1/2} \bigr)  /(3N)$.
Eq-s. \refeq{2.15},
     \refeq{2.16} give sought for time-independent
contributions $\overline{P_1}$ to the
total polarization $P_1(\tau)$ of Eq.
\refeq{2.9} for odd and even numbered spin clusters, respectively.

It remains to find the time-dependent contribution
$P_1^{\sf osc}(\tau)$
to the total polarization $P_1(\tau)$ of
Eq. \refeq{2.6}. Among the four pairs $(I,I')$ in Eq. \refeq{2.6.1},
only the pairs $(I,I')$ with $I \ne I'$ contribute to the
time-dependent part of the function $P_1(\tau)$ of Eq. \refeq{2.6}.
This occurs for $I_B \ge \frac{1}{2}$ only, since otherwise,
i.e. for $I_B = 0$, the allowed values $I = I' = \frac{1}{2}$ are encountered
already in the time-independent polarization $\overline{P_1}$.
Thus, among the four pairs $(I,I')$ in Eq. \refeq{2.6.1} only the two pairs,
i.e. $(I = I_B + \frac{1}{2},I' = I_B - \frac{1}{2})$ and $(I = I_B - \frac{1}{2},I'
= I_B + \frac{1}{2})$ are allowed and provide complex conjugate
contributions to the real-valued function $P_1^{\sf osc}(\tau)$.
It suffices to deal with the first pair, $(I = I_B + \frac{1}{2},I' = I_B - \frac{1}{2})$.
The polarization becomes
\bea{2.17}
\lefteqn{ P_1^{\sf osc}(\tau) =
              2^{-(N_B-1)}
              \sum\limits_{ I_B = \frac{1}{2} }^{N_B/2}
               w(I_B)
\sum\limits_{ m  = -( I_B + \frac{1}{2} )}^{I_B + \frac{1}{2}  }
\sum\limits_{ m' = -( I_B - \frac{1}{2} )}^{I_B - \frac{1}{2}  }
                \delta_{m, m'}  \times
        }                                    \nonumber      \\
&                  &
 \times 2 \cos( 2\tau (I_B + \frac{1}{2} ) )
 \Bigl(
      \sum\limits_{ m_A = \pm \frac{1}{2} }
       m_A  C_{\frac{1}{2}, m_A; I_B, m'-m_A }^{ I_B - \frac{1}{2}, m' }
            C_{\frac{1}{2}, m_A; I_B, m-m_A }^{ I_B + \frac{1}{2}, m }
\Bigr)^2   .
\eea
To complete the derivation of the function $P_1^{\sf osc}(\tau)$,
we use the expression for the
factor
$w(I_B)$ of Eq. \refeq{2.5.2},
the C-G coefficients of Eq. \refeq{2.5.01},
and to sum in  Eq. \refeq{2.17} over the variables $m$ and $m'$ for the fixed value of $I_B$,
yielding
\bea{2.18}
P_1^{\sf osc}(\tau) =
\frac{ 2^{-N_B + 3} }{3(N_B + 1)}
      \sum\limits_{ I_B =  \frac{1}{2} }^{N_B/2}
      {N_B + 1  \choose \frac{1}{2}N_B + I_B + 1 }
            I_B(I_B + 1)
      \cos( 2\tau (I_B + \frac{1}{2} ) ) .
\eea
Finally, by gathering the expressions for the $\overline{P_1}$ in  Eq-s.  \refeq{2.15},
\refeq{2.16} and  the expression $P_1^{\sf osc}(\tau)$ in  Eq. \refeq{2.18}
we'll  go over
(with the substitution $k = I_B-\frac{1}{2}$ for even $N$ and $k = I_B$ for odd $N$)
to the total polarization on the first spin
\bea{2.19}
P_1(\tau) =
\frac{N+2-2^{1-N} { N \choose N/2 } }{3N}
+
\frac{2^{4-N}}{3N}
\sum\limits_{k=0}^{\frac{N}{2} -1 }
A_k(N) \,
\cos( \tau (N - 2k ) )
\eea
for an even $N$-cluster, and
\bea{2.20}
P_1(\tau) =
\frac{N+2}{3N}
+
\frac{2^{4-N}}{3N}
\sum\limits_{k=0}^{\frac{N-1}{2} }
A_k(N) \,
\cos( \tau (N - 2k ) )
\eea
for an odd $N$-cluster, with coefficient
\bea{2.20.1}
A_k(N) = \Bigl( \frac{N+1}{2} - k \Bigr)
         \Bigl( \frac{N-1}{2} - k \Bigr)
{ N \choose k }   \nonumber
\eea
holding in both cases. The formulas
\refeq{2.19}, \refeq{2.20} are
the central result of the paper, and they can accurately
describe a variety of systems in the next section.


\noindent
\section{ Discussion}
\nopagebreak
\vskip 5mm
\subsection{ Non-Ergodic Spin Dynamics }

As Eq. \refeq{2.20} states, for large odd $N$-clusters, the
time average polarization $\langle P_1(\tau) \rangle$ of the spin $1$
tends to $\frac{1}{3}$, while the time average polarization $\frac{2}{3N}$
of any other spin tends to $0$, i.e. polarization of the spin $1$
does not spread uniformly over the $N$-spin
cluster. We call this behavior as non-ergodic spin dynamics
to confront it with the ergodic spin dynamics providing  $1/N$
polarization for all spins in the $N$-spin ensemble.
Figure $2(A)$ shows the behavior of the polarization
$P_1(\tau)$
for the series of odd $N$-clusters.
The principle features of the periodic pulses of
the polarization are managed by two factors:
the time reversibility of the dynamics affects the
exact re-entrance of the polarization to the prepared value
$ P_1(0) = 1 $
after each period $4\pi/g$,
and, the second,
gives rise to the temporary interval
with
the time-independent polarization of the spin $1$, see Appendix for details.
For large $N$-clusters, the total period $4\pi/g$ can be partitioned into the
switching time
$
t^{\sf sw} = 4\pi \frac{{\cal O} (1 )}{ g \sqrt{N} }
$
and the stopping time
$t^{\sf st} = \frac{4\pi}{g} (1 - \frac{ {\cal O} (1 ) }{ \sqrt{N} } )$;
recall $\tau = \frac{1}{2} gt $.
By referring to the Appendix,
the polarization $P_1(\tau)$ is
peaked at the moments $t = 0, 2\pi/g, 4\pi/g, \dots $. The profile of the function
$P_1(\tau)$ around the moment $\tau = 0$  is
\bea{3.1.0}
P_1(\tau) = \frac{1}{3} + \frac{2}{3}\Bigl( 1 - \tau^2 N
\Bigr)
\exp
\Bigl( - \tau^2 N /2
\Bigr).
\eea
The same profile of the function $P_1(\tau)$ holds around all the moments $\tau = m\pi$, for all integer m.
The interval
between the successive peaks and their width are
\bea{3.1.1}
T = \frac{2\pi}{g}    \quad \mbox{and} \quad
\Delta_T = 4 \pi \frac{ {\cal O} (1)}{ g\sqrt{N} },
\eea
respectively.
In other words, for large $N$-clusters, $N \gg 1$,
almost all the time
the polarization of the spin $1$ stays at the fixed value
$\overline{P_1} = 1/3$.
The oscillating part of the function $P_1(\tau)$ is the odd function
of the time with respect to the
moments
$\tau = \frac{\pi}{2}, \frac{3\pi}{2}, \ldots$,
as it is apparent from the $P_1(\tau)$ of Eq. \refeq{2.20}.

Figure $2(B)$  shows the profiles of the
polarization for even $N$-spin clusters.
For large even values of $N$,
the polarization at spin $1$ stays fixed over the
long time interval
$
t^{\sf st} = \frac{2\pi}{g}(1  - \frac{ {\cal O} ( 1 ) }{ \sqrt{N} } )
$
within each period $2\pi/g$. Unlike odd $N$-clusters, the profiles
$P_1(\tau)$ for even $N$-clusters are even functions of the time
with respect to the time moments
$\tau =\pi, 2\pi, \ldots$.

Relying on
the experimental time interval $T$ and the width of the pulses
$\Delta_T$, the two relations in Eq. \refeq{3.1.1}
together with the expressions for the coupling $g$ of Eq.-s \refeq{1.21}, \refeq{1.24}
give
the volume and form-factor (the angle $\alpha$ is assumed to be known)
\bea{3.1.2}
V = \frac{4}{c}
\Bigl( \frac{T}{\Delta_T} \Bigr)^2
\quad \mbox{and} \quad
f(\frac{a}{b}) P_2(\cos\alpha) =
\frac{8}{c} \frac{T}{\gamma^2 \hbar\Delta_T^2},
\eea
where $c=N/V$ denote the concentration of the molecules carrying
the spin-$\frac{1}{2}$.


\noindent
\vskip 5mm
\subsection{Polarization Dynamics in Fluctuating Nano-bubbles}

The Eq.-s \refeq{2.19}, \refeq{2.20} can be adopted to accounting
for a time dependence of the volume of the nano-cavity and, thus,
providing a means to explore NMR imaging of the cavitation bubbles
in water \cite{Bren}, blood \cite{Bru}, etc., alongside with the
conventional high-speed photography.
Dynamics of the surface of a typical bubbling behavior occurs on
a millisecond time scale \cite{Bren}, i.e. on the same time scale
which is relevant for the nuclear spin dynamics. It is legitimate
therefore to question here: how does the dynamics
of the nano-sized  volume  affect the nuclear spin dynamics ?
Our intention in this section is to show that the fluctuations of the nano-volume
(governed either by an external inputs or by inherently thermal noise)
drive the polarization to the non-ergodic value $1/3$
irreversibly so that the periodic in time pulsating of
the polarization break down as the time proceeds.

The formulation of Section \ref{sec. 2} is easily extended to the
case of
time varying volume $V$ since the coupling $g(V(t))$ enters into the Hamiltonian \refeq{2.1}
as a common factor in front of the whole operator part.
The functional form  of the polarization $P_1^{\sf osc}(\tau)$ of Eq. \refeq{2.18}
which has been derived for time independent coupling $g$ is
generalized to the case of the
function $g(t)$ provided that the time $\tau = gt/2$ in Eq.  \refeq{2.18}
is now replaced with a new time
\bea{3.3.1}
\tau =
\frac{1}{2}gt \longrightarrow \frac{1}{2}\int_0^t g(t') dt' .
\eea
We will be interested in the transformation \refeq{3.3.1}
\bea{3.3.2}
g(t) = \langle g \rangle  +   \delta g (t),
\eea
where the $\delta g (t)$ stands for the
Gaussian random noise characterized by the first two moments
\bea{3.3.3}
\langle \delta g (t) \rangle = 0, \quad
\langle \delta g (t_1) \delta g (t_2)  \rangle =
\langle ( \delta g )^{2}  \rangle  \,  \gamma( | t_1 - t_2 |),
\eea
where
$\langle ( \delta g )^{2}  \rangle$
is the variance and the $\gamma(t)$ denotes the correlation function,
for example,
$
\gamma(t) = \exp(-t/t_{\sf c})
$,
with $t_{\sf c}$ being the correlation time.
By the comment before Eq. \refeq{3.3.1}, we replace the factor
$
\cos \left( 2\tau (I_B + \frac{1}{2} ) \right)
$
in Eq. \refeq{2.18} with expression
$
\cos  \left[ (I_B + \frac{1}{2}) \left( \langle g \rangle t +
                                        \int_0^t \delta g(t') dt'
                                 \right)
      \right]
$.
The Gaussian averaging of this factor over the random function $\delta g (t)$ is carried out
by the formula, see e.g. \cite{Abragam},
\bea{3.3.3}
\left \langle
\exp \left( i (I_B + \frac{1}{2} )  \int_0^t \delta g(t') dt' \right)
\right\rangle_{\delta g}
=
\exp \Bigl(
- ( I_B + \frac{1}{2})^2
\langle ( \delta g )^{2}  \rangle  \,
 T^2
     \Bigr),
\eea
with
\bea{3.3.4}
 T^2 =
\int_0^t (t- t') \,  \gamma(t')\, dt'
     .
\eea
We confine our attention, first, with the
polarization  for even $N$, with  $N\gg 1$, and then close the
section with the final result for odd $N$, $N\gg 1$.
Let us write the polarization of Eq. \refeq{2.18} with the averaging
Eq. \refeq{3.3.3}  as
\bea{3.3.5}
 P_1 (t) &  = &  \overline{P_1} +
                  \frac{ 2^{-N_B + 3} }{3(N_B + 1)}
                  \sum\limits_{ I_B = \frac{1}{2} }^{N_B/2}
                  {N_B + 1  \choose \frac{1}{2}N_B + I_B + 1 }
                   I_B (I_B + 1) \Phi(t),
                                     \nonumber          \\
 \Phi(t) &  =  &
          \exp
          \left(
           - \left( I_B + \frac{1}{2} \right)^2
             \langle ( \delta g )^{2}  \rangle \,
          T^2
         \right)
                        \cos\left( \langle g \rangle  t (I_B +
                \frac{1}{2})\right)
                                .
\eea
The exponent in Eq. \refeq{3.3.5} tell us that the successive peaks
of time dependent part of the function $P_1(t)$,
reduce  to zero
at  $t \to \infty $,
thus, only the
time independent part, i.e. $\overline{P_1} = 1/3$
of Eq. \refeq{2.14.1} of the
function  $P_1(t)$ survives at  $t \to \infty $,
after the Gaussian averaging over the function $\delta g(t)$.
The integral over $t'$ in the constant $T^2$  of Eq. \refeq{3.3.4}
can be evaluated in the two asymptotic cases, for the large
and the small correlation time \cite{Abragam},
\be{3.3.8}
T^2 =
    \cases{
  t^2/2      ,  & if  $ \quad    t_{\sf c}^2 \, \langle ( \delta g )^{2}  \rangle \gg  1  $  \cr
 t_{\sf c} t ,  & if  $ \quad    t_{\sf c}^2  \, \langle ( \delta g )^{2}  \rangle \ll  1  $  \cr
          }
   \quad \mbox{,}
\ee
In order to find the function $P_1(t)$ of Eq. \refeq{3.3.5} at
asymptotics  $N \gg 1$, we can
replace the sum in Eq. \refeq{3.3.5} with the Gaussian averaging,
just as done in Eq. (A.3) of the Appendix, yielding
\bea{3.3.6}
P_1(t) =
\frac{1}{3} +
\frac{16}{ 3N^{3/2} \sqrt{\pi/2} } \,
\sum\limits_{ n = 1 }^{ N/2 }
\cos \left(  \langle g \rangle t n \right)
     \left( n^2 - \frac{1}{4} \right)
e^{-a n^2} ,
\eea
where
\bea{3.3.7}
a = \frac{2}{N}  +
\langle (\delta g)^2  \rangle
   T^2 .
\eea
Figure $3$ shows the polarization dynamics of a single spin within $N=134$
spin aggregate
for
$\langle (\delta g)^2  \rangle / \langle  g  \rangle^2 = 10^{-4}$.
Based upon the Appendix, the sum over $n$ in Eq. \refeq{3.3.6}
is simplified with the Poisson's
summation formula. On defining the partial sums entering into Eq. \refeq{3.3.6} by
\bea{3.3.9}
S_1(t) & = & \sum\limits_{ n = 1 }^{N/2}
\cos \left( \langle g  \rangle  t n \right) e^{- a n^2 } = \nonumber \\
       & = &
- \frac{1}{2 } +
  \frac{1}{2 }
\sqrt{ \frac{\pi}{a} }
\sum\limits_{ q = - \infty }^{ \infty }
\exp
        \Bigl(
- \pi^2 \Bigl( q + \frac{ \langle g \rangle  t}{2\pi} \Bigr)^2 / a
        \Bigr),
\eea
and
\bea{3.3.10}
  &S_2(t)  =  \sum\limits_{ n = 1 }^{N/2}
\cos \left( \langle  g \rangle  t n \right)  n^2 e^{- a n^2 } =  -\partial
 S_1(t)/\partial a  =  & \nonumber \\
 &   =\frac{ \sqrt{\pi} }{ 4 a^{3/2} }
\sum\limits_{ q = - \infty }^{ \infty }
\Bigl( 1 - \frac{2}{a}
           \pi^2 \Bigl( q + \frac{ \langle g \rangle  t}{2\pi} \Bigr)^2
\Bigr)
\exp
        \Bigl(
- \pi^2 \Bigl( q + \frac{ \langle g \rangle  t}{2\pi} \Bigr)^2 / a
        \Bigr), &
\eea
we get
\bea{3.3.11}
P_1 (t) = \frac{1}{3} +
\frac{ 16 }{ 3  N^{3/2} \sqrt{ \pi/2 } }
\left( S_2(t) -   \frac{1}{4} S_1(t) \right).
\eea
To find the envelope of the successive peaks of the function
$P_1(t)$ of Eq. \refeq{3.3.11}, put into Eq.-s \refeq{3.3.9} - \refeq{3.3.11} the time
$t = 2\pi m / \langle g \rangle $,
with  $m$ running over integer numbers.
This gives the polarization at the
discrete moments $m$,
\bea{3.3.12}
S_1(m)  =  \sum\limits_{ n = 1 }^{N/2}  e^{- a n^2 } =
- \frac{1}{2 } +
  \frac{1}{2 }
\sqrt{ \frac{\pi}{a} }
\sum\limits_{ k = - \infty }^{ \infty }
e^{ - \pi^2 k^2 / a },
\eea
\bea{3.3.13}
S_2(m) =  \sum\limits_{ n = 1 }^{N/2}
 n^2 e^{- a n^2 } =
   \frac{ \sqrt{\pi} }{ 4 a^{3/2} }
\sum\limits_{ k = - \infty }^{ \infty }
\Bigl( 1 - \frac{2}{a}
           \pi^2 k^2
\Bigr)
e^{- \pi^2 k^2 / a }.
\eea
The functions $S_1(m)$ and $S_2(m)$
inherit their dependence on the "time" $m$ through the constant  $a$
 of Eq.-s \refeq{3.3.7}, \refeq{3.3.8}
\be{3.3.14}
a = \cases{
\frac{2}{N} +
2 \pi^2 m^2
 \langle (\delta g)^2 \rangle
    / \langle  g  \rangle^2   , & if
$\quad   t_{\sf c}^2 \langle ( \delta g )^{2}  \rangle \gg  1 $ \cr
\frac{2}{N} +
2 \pi m
 t_{\sf c} \langle (\delta g)^2 \rangle /  \langle  g  \rangle ,           & if
$\quad  t_{\sf c}^2 \langle ( \delta g )^{2}  \rangle \ll  1   $ \cr
          }
  \quad \mbox{,}
\ee
where we substitute $t = 2\pi m /\langle  g  \rangle $ into
Eq.-s \refeq{3.3.7}, \refeq{3.3.8}. For $N \gg 1$ and $m \gg 1$, we
 drop the summand $2/N$ in Eq. \refeq{3.3.14} assuming that
 $a \gg 1$,
\be{3.3.15}
a  \approx  \cases {
2 \pi^2 m^2
 \langle (\delta g)^2 \rangle
    / \langle  g  \rangle^2
, & if
$ \quad   t_{\sf c}^2 \langle ( \delta g )^{2}  \rangle \gg  1 $ \cr
2 \pi m
 t_{\sf c} \langle (\delta g)^2 \rangle /  \langle  g  \rangle
, & if
$ \quad   t_{\sf c}^2 \langle ( \delta g )^{2}  \rangle \ll  1 $ \cr
           }
   \quad \mbox{.}
\ee
For $a \gg 1$, we find the sums over the $k$ in
Eq. \refeq{3.3.12} and Eq. \refeq{3.3.13} by making use, again,
the Poisson's summation formula, Eq. (A.5) of the Appendix,
that accelerates the convergence of the sums for $a \gg 1$.
On reading Eq. (A.5) of the Appendix backward from the r.-h. side to the l.-h. side, we get
\bea{3.3.16}
I(a)  =
\sqrt{ \frac{\pi}{a} }
\sum\limits_{ k = - \infty }^{ \infty }
e^{ - \pi^2 k^2 / a } =
\sum\limits_{ \ell = - \infty }^{ \infty }
e^{ - a \ell^2 } = 1 + 2e^{ - a } + {\cal O}(e^{ - 4a } ).
\eea
Thus, Eq.-s \refeq{3.3.9}, \refeq{3.3.10} become
\bea{3.3.17}
S_1(m)  =  - \frac{1}{2 } + \frac{1}{2 } I(a) = e^{ - a },
\eea
\bea{3.3.18}
S_2(m) = - \frac{1}{2} \frac{\partial I(a)}{\partial a} = e^{ - a },
\eea
so that Eq. \refeq{3.3.11} gives the
 polarization of the first spin
\bea{3.3.19}
P_1(m) = \frac{1}{3} +
\frac{ 4 \sqrt{2} }{ N^{3/2} \sqrt{ \pi } }
e^{ - a }
\eea
with the $a$ from Eq. \refeq{3.3.15}.
Eq. \refeq{3.3.19} implies the total polarization of the first spin
\bea{3.3.20}
P_1(m) = \frac{1}{3} +
\frac{ 4 \sqrt{2} }{ N^{3/2} \sqrt{ \pi } }
e^{ - a }.
\eea
We conclude this section with the result for the total polarization
for odd total number, $N$,  of spins.
Due to alternating peaks of the polarization $P_1(t)$ of Eq. \refeq{2.20},
see also Figure 2A, we get
\be{3.3.21}
P_1(m)  = \cases{
\frac{1}{3} -
\frac{ 4 \sqrt{2} }{ N^{3/2} \sqrt{ \pi } }e^{ - a },  &
$\quad \mbox{ for large odd m} $ \cr
\frac{1}{3} +
\frac{ 4 \sqrt{2} }{ N^{3/2} \sqrt{ \pi } }e^{ - a },  &
$ \quad \mbox{ for large even m} $ \cr
                 }
   \quad \mbox{.}
\ee
Eq. \refeq{3.3.21} with the $a$ from Eq. \refeq{3.3.15} shows that
the the polarization peaks, $P_1(m)$,   of
spin-carrying gas has the
Gaussian and the exponential time dependence  for the
large and the small
correlation times of the fluctuations of the nano-bubbles, respectively.


\noindent
\vskip 5mm
\subsection{NMR Line Shape}

To calculate the NMR line shape exactly, we use the same effective
Hamiltonian of Eq. \refeq{2.1} as described in Section \ref{sec. 2}.
The NMR line shape is the Fourier transform of the free induction
decay (FID), $F(t)$, of the $N$-spin ensemble \cite{Abragam}.
It is the NMR line
shape on the protons in hydrogenated thin silicon
films  that got the first experimental
evidence for the validity of the effective
Hamiltonian of Eq. \refeq{2.1}
in the nano-cavities \cite{Baugh}. We are interested in the FID signal
\bea{3.4.1}
F(t) =
tr
\left.
\left\{
 e^{iHt} I_{+} e^{-iHt} I_{-}
\right\}
\right/
tr
\left.
\left\{
I_{+}I_{-}
\right\}\right. ,
\eea
with
$ I_{\alpha} = \sum_{n=1}^{N}  I_{n\alpha}$,
$ I_{\pm} = I_{x} \pm i I_{y}$, $\alpha = x,y,z$. The reason for
an exact solution for the
FID of Eq. \refeq{3.4.1} is that
the total Hamiltonian of Eq. \refeq{2.1} can be expressed in terms of
the $3$ collective spin operators $I_{\alpha}$, instead of $3N$
spin operators in $H$ of Eq. \refeq{2.1}. Due to commutation relations
$[{\mathbf I}^2, I_{\alpha}]=0$, rewrite Eq. \refeq{3.4.1} as
\bea{3.4.2}
F(t) =
tr
\left.
\left\{
 e^{iGtI_z^2} I_{+} e^{-iGtI_z^2} I_{-}
\right\}
\right/
tr
\left.\left\{
I_{+}I_{-}
\right\}\right. ,
\eea
with $G = 3g/2$ for dipolar interactions in the effective Hamiltonian of Eq.
\refeq{2.1} with $\zeta = 2$. The Heisenberg equations of motion for operator
$
I_{+}(t) = e^{iGtI_z^2} I_{+} e^{-iGtI_z^2}
$
is
solved exactly to
$I_{+}(t) = \exp( iGt(2I_z - 1) )I_{+}(0)$, $I_{+}(0) = I_{+}$.
The averaging in Eq. \refeq{3.4.1} is performed in
$N !/( N_{\uparrow}! N_{\downarrow}!)$-fold degenerate
basis of the states ($N_{\uparrow}, N_{\downarrow}$)
with $N_{\uparrow}$, ($N_{\downarrow}$) spins
up (down), so that
$N_{\uparrow} + N_{\downarrow} = N$ and
$ I_z = ( N_{\uparrow} - N_{\downarrow} )/2$.
The averaging gives the FID
\bea{3.4.3}
F(t) = \left[ \cos \left(\frac{3}{2}gt\right)   \right]^{N-1}.
\eea
The effect of dephasing of the proton spins within the nano-cavity due to
the interactions with the protons at the surface of the nano-cavity is
 introduced
phenomenologically as
\bea{3.4.3.1}
F(t) = \left[ \cos \left(\frac{3}{2}gt\right) \right]^{N-1} e^{-t/T_2}.
\eea
where, the time $T_2$ relevant for the experiments
\cite{Baugh} is $T_2 \approx 1 \div 3$ ms.
The moments of the line shape are
\bea{3.4.4}
M_n = \int_{-\infty}^{\infty} d\omega \omega^n \Im(\omega)  =
       \left.\left[ \frac{d^n F(t)}{d(it)^n} \right] \right|_{t=0}.
\eea
where $\Im(\omega)$ enters through the Fourier transformation of the FID,
\bea{3.4.5}
F(t) = \int_{-\infty}^{\infty} d\omega \Im(\omega) e^{i\omega t}.
\eea
The second and the fourth order moments are inferred from Eq.
\refeq{3.4.3}, as
\bea{3.4.5}
M_2 = (N-1)\left(3g/2\right)^2, \quad  M_4 = (N-1)(3N-5)\left(3g/2\right)^4 .
\eea
The $M_2$ derived in \cite{Baugh} by the Van-Fleck formula coincides with
the $M_2$ of Eq. \refeq{3.4.5}, as it should be. The line shape in the nano-cavity
volume appears to be the volume dependent allowing for
determining the volume of the nano-pores in
hydrogenated silicon films \cite{Baugh}.


\noindent
\section{Conclusion}
\label{sec. 4}

We have presented the exact
time-dependent description
of spin-$\frac{1}{2}$ dynamics with infinite
range spin interactions and the
initial polarization prepared on a single spin $1$, i.e. $P_1(0)=1$.
Spin dynamics for odd and even
numbered clusters demonstrates the periodic pulses of
the polarization $P_1(\tau)$ on the spin $1$.
For  large odd $N$-clusters,
the polarization on the spin $1$ has pulses
over the time interval
$
t^{\sf sw} = 4\pi \frac{{\cal O} (1 )}{ g \sqrt{N} }
$,
from $P_1(0)=1$
to the time-independent polarization which lasts, thus,
$t^{\sf st} = \frac{4\pi}{g} (1 - \frac{ {\cal O} (1 ) }{ \sqrt{N} } )$
within any period $4\pi/g$.
For large even $N$-clusters, the switching time
$
t^{\sf sw} = 2\pi \frac{{\cal O} (1 )}{ g \sqrt{N} }
$,
and the period equals
$2\pi/g$.
The stationary polarization on the spin $1$ is non-ergodic,
as its value tends to $1/3$
(instead of tending to the ergodic value $1/N$)
as $N$ tends to infinity.
The profiles of the polarizations within the
series of odd (even) large clusters are remarkably similar.

The specific polarization profile in the clusters with
infinite range spin interactions is in a sharp contrast
to the profiles of the polarization in $1D$ clusters with
nearest neighbor XY Hamiltonian \cite{FBE}. Two differences can be drawn
from the presented results:

(1) The overall behavior of the polarization $P_1(t)$
in the system with infinite range interaction
is strictly reversible,
periodic with the period $4\pi/g$ for any $N$,
whereas, on a large $1D$  chains, $N\gg 1$, with XY Hamiltonian
the polarization $P_1(t)$ on the spin $1$
moves  in irregular fashion.

(2) For large $N$-spin clusters, $N \gg 1$,
the polarization $P_1(t)$ of the spin $1$
exhibits the plateau region
at non-ergodic value $\overline{P_1} = 1/3$
and the pulses of the polarization $P_1(t)$
have a short time span of about
$
4\pi {\cal O}(1 )/( g \sqrt{N})
$.
This is in
contrast to the behavior of the
polarization $P_1(t)$ in $1D$ spin chains with XY Hamiltonian
where polarization on the spin $1$
depends on the time in irregular fashion with $t^{\sf st} = 0$.

Finally, the paper demonstrates  the sensitivity of the polarization dynamics
( reversibility and ergodicity in the many-spin systems)
to the radius of interaction.
Incorporation into the theory of the real dipolar interactions
is the most challenging task of dynamical theory
and the accurate answer is not settled yet,
although the general picture of the spin dynamics is known to
be diffusional  \cite{ZC}.


\vskip 5mm

\noindent

{\bf Acknowledgments}

\vskip 5mm

Thanks are due
to D.E. Fel'dman and S.V. Iordanskii for helpful discussions,
to A.K. Khitrin for sending the report \cite{Baugh},
to S.I. Doronin and I.I. Maximov for the help in preparing the manuscript  and
to the RFBR for funding under the Grant No. 01-03-33273.


\vskip 5mm

\noindent

{\bf Appendix. Derivation of Eq. \refeq{3.1.0} }

\vskip 5mm

We want to prove that the function
$P_1^{\sf osc}(\tau)$
in Eq. \refeq{2.18}
for $N\gg 1$ has the form of pulses of width
$$
\Delta_T =  4 \pi \frac{ {\cal O} (1 )}{ g\sqrt{N} }
$$
at equidistant time moments $\tau = 0, \, 2\pi, \, 4\pi, \ldots $
so that the profile of the function $P_1^{\sf osc}(\tau)$
at the time
moment $\tau = 0$ is
$$
P_1^{\sf osc}(\tau) =
\frac{2}{3}\Bigl( 1 - \tau^2 N
\Bigr)
\exp
\Bigl( - \tau^2 N /2
\Bigr), \quad \mbox{for $N\gg 1$}
\eqno (A.1)
$$
To prove Eq. (A.1),
we introduce a new variable
$n = I_B + \frac{1}{2}$
in the expression $P_1^{\sf osc}(\tau)$ of Eq. \refeq{2.18}, so that the function $P_1^{\sf osc}(\tau)$
takes the form ( recall that the total number of spins equals $N = N_B + 1$)
$$
P_1^{\sf osc}(\tau) =
\frac{16}{ 3N }
\sum\limits_{ n = 1 }^{N/2}
 2^{-N}
 {N  \choose \frac{1}{2}N + n }
            \Bigl(
            n^2 - \frac{1}{4 }
            \Bigr)
      \cos\left( 2\tau n  \right) .
\eqno (A.2)
$$
Next, we use asymptotic formula for
the binomial coefficient,
$$
 2^{-N}
 {N  \choose \frac{1}{2}N + n } =
\frac{ 1 }{ \sqrt{\pi N /2 }  } \exp
                    \left(
                    - \frac{ n^2 }{ N/2 }
                    \right)
                    \left(
         1 + n^3 \frac{   {\cal O }(1)  }{ \sqrt{N}  }
                    \right).
\eqno (A.3)
$$
Eq. (A.3) allows one to look at the summation in  Eq. (A2)
as an averaging  over the Gaussian distribution function.
To simplify the calculations of Eq. (A2) further, we apply the Poisson's identity
\cite{MoFe}
$$
\sum\limits_{ \ell = - \infty }^{ \infty }
\cos\left( 2\pi \epsilon \ell \right) e^{-a \ell^2} =
\sqrt{ \frac{\pi}{a} }
\sum\limits_{ k = - \infty }^{ \infty }
e^{- \pi^2 (k + \epsilon)^2 /a }.
\eqno (A.4)
$$
In many circumstances,
including our ones, the resulting sum over the $k$ in the r.-h.
side of  Eq. (A.4) converges much
faster than the original sum over the $\ell$ in the l.-h. side of  Eq. (A.4).
To apply  Eq.-s (A.4)  to Eq. (A.2), we expand the sum in Eq. (A.2)
up to $n = \infty$ since the terms of
the sum in Eq. (A.2) practically vanish for $n > N/2$ and  $N \gg 1$.
Thus, by the Poisson's identity of Eq. (A.4), we introduce the sum
( a partial contribution to the sum of Eq. (A.2) )
$$
S_1(\tau) =
\sum\limits_{ n = 1 }^{N/2}
\cos\left( 2\tau n \right) e^{- n^2 /(N/2) } =
- \frac{1}{2 } +
  \frac{1}{2 }
\sqrt{ \frac{\pi N}{2} }
\sum\limits_{ k = - \infty }^{ \infty }
\exp
        \Bigl(
- \pi^2 \Bigl( k + \frac{\tau}{\pi} \Bigr)^2  \frac{N}{2}
        \Bigr).
\eqno (A.5)
$$
In order to show that the function $S_1(\tau)$ has the form of the Gaussian peaks at
equidistant moments $\tau = 0,\pm \pi, \pm 2\pi, \ldots$
it suffices  to analyze the function
$S_1(\tau)$  around the point $\tau =0$.
In this case, the leading contribution to the sum in Eq. (A.5) is provided
by the term $k=0$. Denote that  if we analyze the peak around $\tau = m \pi$, where $m$ is
integer,
then the leading contribution to the sum
$S_1(\tau)$ in Eq. (A.5)
comes from the $k= -m$.
Thus, in considering $N \gg 1$, we can drop all the terms in Eq. (A.5), except the leading term
$k=0$, yielding
$$
S_1(\tau) =
- \frac{1}{2 } +
\frac{1}{2 }
\sqrt{ \frac{\pi N}{2} }
e^{ - \tau^2 N/2 } .
\eqno (A.6)
$$
Analogously, we determine the partial sum
$$
\begin{array}{clc}
S_2(\tau) & = &
\sum\limits_{ n = 1 }^{(N - 1)/2}
\cos\left( 2\tau n \right) n^2 e^{- n^2 /(N/2) }  =  \\
          & = & -  \frac{ \partial}{  \partial (2/N) } S_1(\tau)  =
\frac{ N^{3/2} \sqrt{\pi} }{ 8 \sqrt{2}}
\left( 1 - \tau^2 N \right)
e^{ - \tau^2 N /2 }  .
\end{array}
\eqno (A.7)
$$
At $ N\gg 1$,  the function $S_1(\tau)$ of Eq. (A.6)
has a negligible contribution to the function
$$
P_1^{\sf osc}(\tau) =
\frac{ 16 }{ 3  N^{3/2} \sqrt{ \pi/2 } }
\left( S_2(\tau) -   \frac{1}{4} S_1(\tau) \right)
$$
in comparison with the contribution of the function $S_2(\tau)$ of Eq. (A.7),
yielding the sought for result of Eq. (A.1).

In general, the function $P_1^{\sf osc}(\tau)$ for an arbitrary
$\tau$  has the pulses at the moments $\tau = k \pi$ with integer
$k$,
$$
P_1^{\sf osc}(\tau) =
\sum\limits_{ k = - \infty }^{ \infty }
\frac{2}{3}\Bigl( 1 - \pi^2 \Bigl( k + \frac{\tau}{\pi} \Bigr)^2  N \Bigr)
\exp
            \Bigl(
- \pi^2 \Bigl( k + \frac{\tau}{\pi} \Bigr)^2  \frac{N}{2}
            \Bigr).
\eqno (A.8)
$$



\newpage
\noindent
{\large  \bf
Captions to figures.
}
\vskip 5mm

Fig.$1$.
Schematic representation of  nano-pore with the
molecules carrying  the nuclear
spins  and are in a rapid thermal motion.

Fig.$2$.
On the panel A, the
polarization $P_1(\tau)$ of Eq. \refeq{2.20}
of the first spin is varied
with non-dimensional time $\tau = gt/2$
for the series of an odd total number, $N$, of spins.
The panel B displays
the polarization $P_1(\tau)$ of Eq. \refeq{2.19}
for the series of even $N$.

Fig.$3$.
Polarization dynamics of a single spin, $P_1(t)$,
within the $N=134$ spin aggregate when the volume of the nano-cavity
fluctuates providing with relative variance for the $g$ coupling
equal to $\langle (\delta g)^2  \rangle/\langle  g  \rangle^2  = 10^{-4}$,
see Eq.-s
\refeq{3.3.6}, \refeq{3.3.7}.


\newpage

\begin{figure}[h]
\begin{picture}(10,10)

\includegraphics*[scale=0.5]{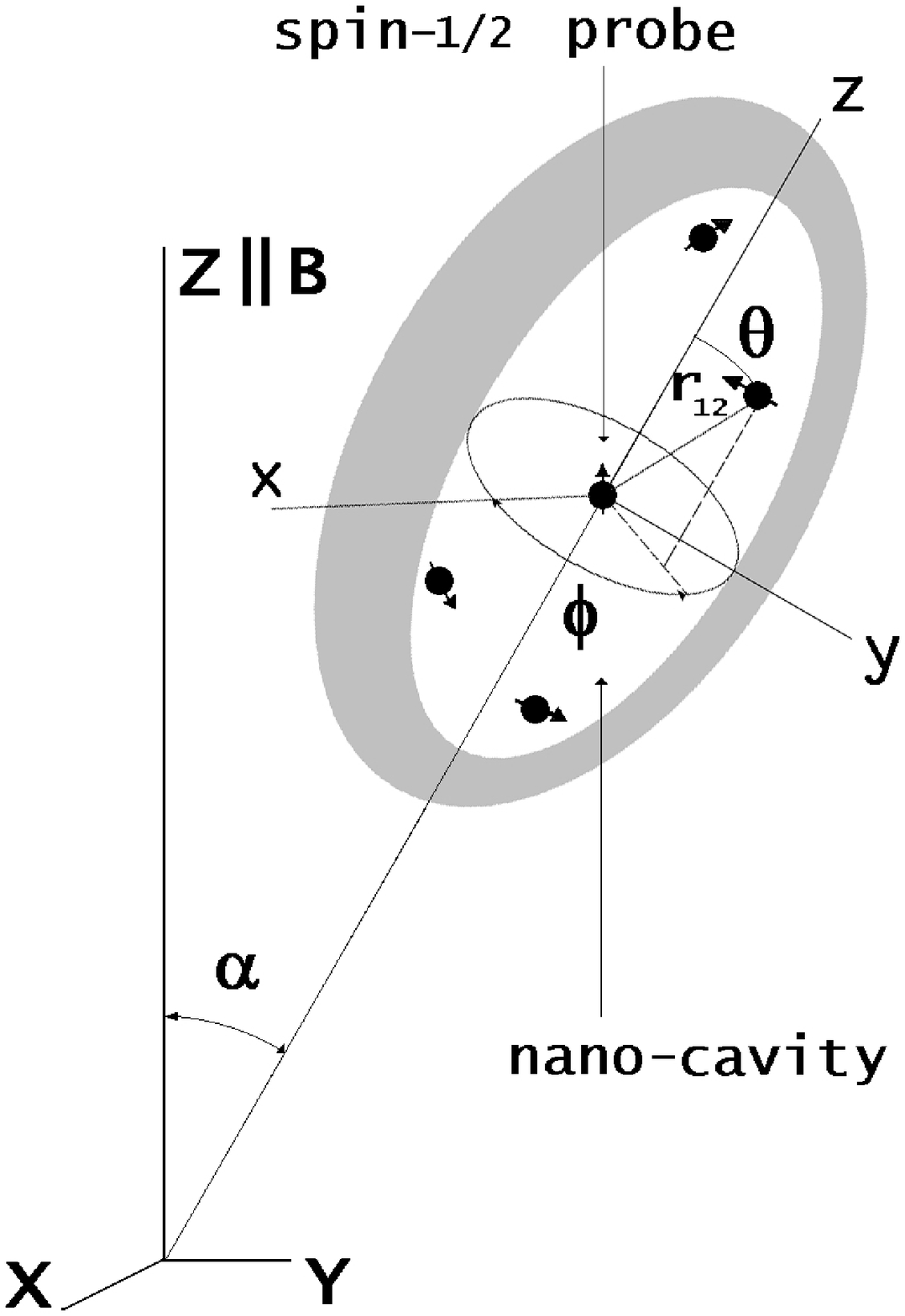}

\end{picture}

\end{figure}


\newpage

\begin{figure}[h]
\begin{picture}(10,10)

\includegraphics*[scale=0.75]{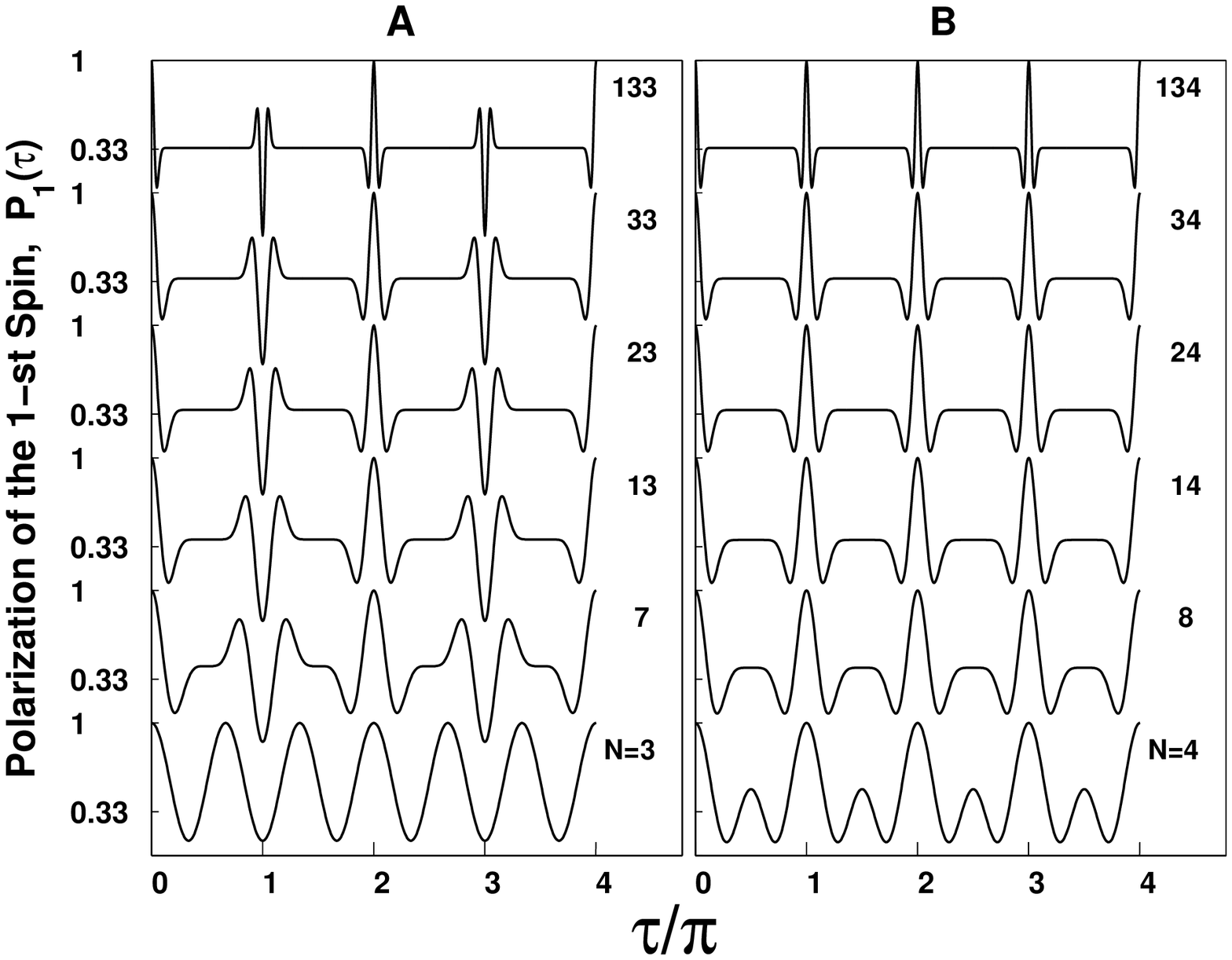}

\end{picture}

\end{figure}


\newpage

\begin{figure}[h]
\begin{picture}(10,10)

\includegraphics*[scale=0.75]{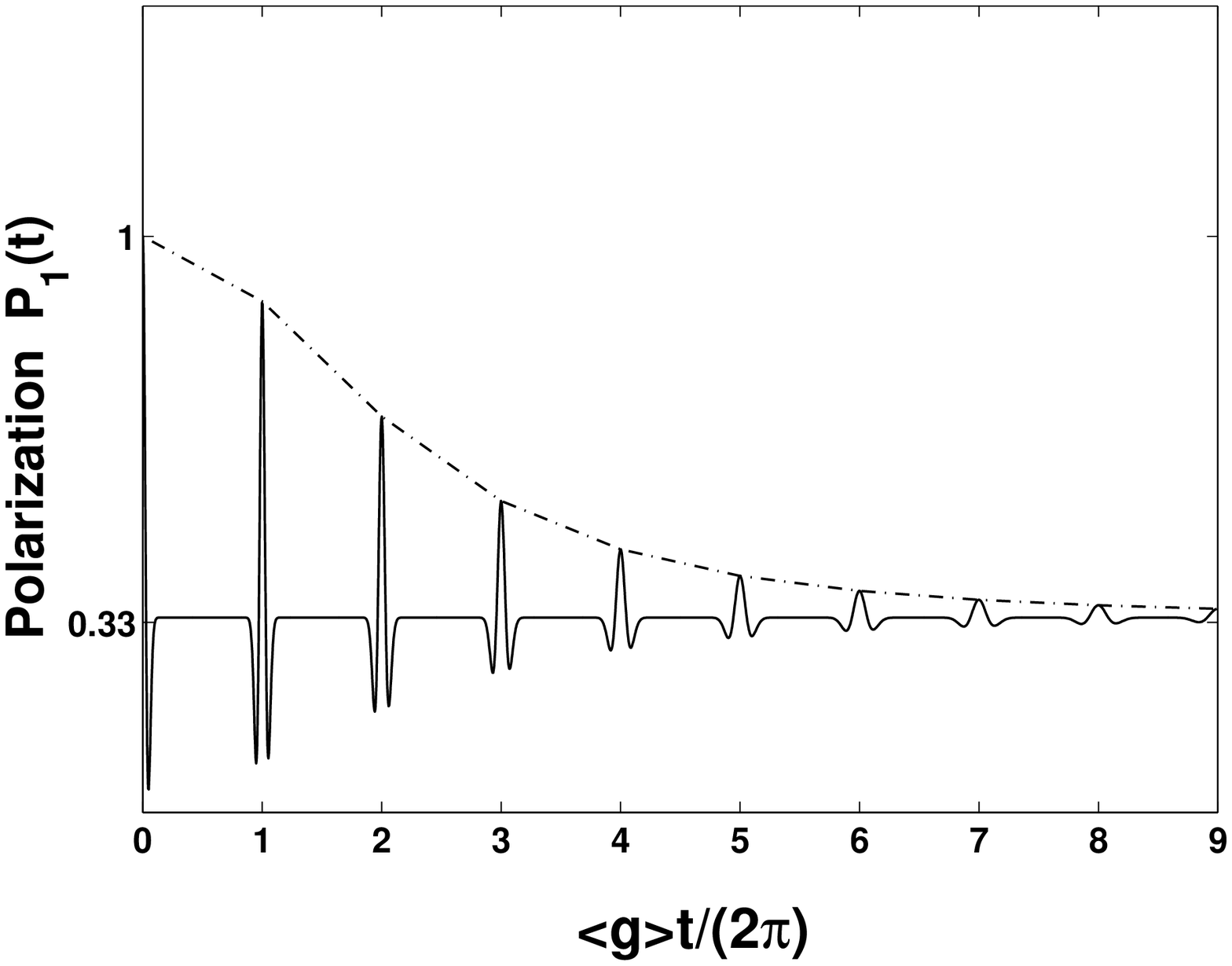}

\end{picture}

\end{figure}


\end{document}